\documentclass[aps,floats,showpacs,amsmath,epsf,twocolumn, superscriptaddress]{revtex4}
\usepackage{amsfonts, relsize, color}
\usepackage{graphics}
\usepackage{graphicx}
\usepackage{subfigure}
\usepackage{hyperref}

\begin{document}

\title{Half vortex and fractional electrical charge in two dimensions}
\author{Bitan Roy}
\affiliation{Condensed Matter Theory Center and Joint Quantum Institute, University of Maryland, College Park, Maryland 20742-4111, USA}

\author{Igor F. Herbut}

\affiliation{ Department of Physics, Simon Fraser University, Burnaby, British Columbia, Canada V5A 1S6}

\date{\today}

\begin{abstract}
Despite fermion doubling, a two-dimensional quasi-relativistic spin-1/2 system can still lead to true fractionalization of electrical charge, when a massive ordered phase supports a ``half-vortex". Such topological defect is possible when the order parameter in form of Dirac mass is described by two $U(1)$ angles, and each of them winds by an angle $\pi$ around a point. We demonstrate that such a mass configuration in an eight-dimensional Dirac Hamiltonian exhibits only a single bound zero mode, and therefore binds the charge of $e/2$. In graphene, for example, such an ordered phase is provided by the easy-plane spin-triplet Kekule valence bond solid. We argue that an application of an in-plane magnetic field can cause an excitonic instability toward such ordered phase, even for weak repulsion, when the on-site, nearest-neighbor and second neighbor components of it are of comparable strengths. 
\end{abstract}


\maketitle

\vspace{10pt}

\section{Introduction}
 
Quantization of physical observables, such as energy, angular momentum, or charge, played the crucial role in the formulation of quantum mechanics, which revolutionized physics over the course of the last hundred years. Successful description of many previously enigmatic experimental observations in solids and in atomic physics established quantum mechanics as the fundamental theory of nature. Despite the ubiquity of quantization, fractionalization of quantum numbers is also possible. It is believed that the existence of real space topological defects~\cite{rajaraman} often, if not always, underlies fractionalization. Celebrated examples of these are the domain wall of a Dirac mass in one dimension~\cite{jackiw-rebbi, schrieffer}, vortex configuration between two Dirac masses with requisite $U(1)$ symmetry in two spatial dimensions~\cite{weinberg, jackiw-rossi}, and t'Hooft-Polyakov monopole and dyon in three dimensions~\cite{jackiw-rebbi}. All these proposals share few common  features: (i) the underlying fermionic dispersion is relativistic in nature, (ii) Dirac fermion becomes massive, with the mass either generated dynamically, or induced externally, (iii) the Dirac mass constitutes a topologically nontrivial background in real space. When these conditions are satisfied~\cite{chiken}, an isolated zero energy state appears at the band center and a fractionalized electrical charge $e/2$ becomes possible.  However, despite many theoretical proposals~\cite{jackiw-rebbi, schrieffer, jackiw-rossi, hou-chamon-mudry, jackiw-pi, herbut-so3, semenoff-jackiw, herbut-cl3, herbut-lu, ryu, roy-axial, roy-smith-kennett, doron}, a realization of fractional electrical charge has remained elusive. The reason is the unavoidable doubling of fermions on a lattice~\cite{ninomiya}, which doubles the number of zero modes as well, and that way restores the standard quantization of electrical charge for spin-1/2 particles such as electrons.

We here demonstrate that despite fermion doubling, when Dirac masses are wound into a specific real space configuration which may be called \emph{half-vortex}, the defect binds exactly one mode at zero energy and leads to a genuine fractionalization of electrical charge in two dimensions. The basic idea goes as follows. Consider an order parameter in a two-dimensional Dirac system represented by a mass term which is described by {\it two} $U(1)$ angles, with each of them winding by $\pi$ around some point [see Eq.~(\ref{SPzeeman})]. While the Hamiltonian remains single-valued, it can be brought into a block-diagonal form,  with one four-component massive block hosting a full vortex (with a twist by an angle of $2 \pi$), while the other block has only a topologically trivial mass term. The first block is then equivalent to the Jackiw-Rossi Hamiltonian~\cite{jackiw-rossi}, and such a half-vortex binds in total only a \emph{single} zero energy state, yielding the electrical charge which is a half of the charge of the underlying constituents to be accumulated near the half-vortex.

In the prototypical two-dimensional Dirac system such as graphene, such order could correspond to the easy-plane components of the spin-triplet Kekule valence bond solid. In this phase, the nearest-neighbor hopping amplitude acquires a commensurate periodic modulation, which, however, has the opposite sign for two projections of electron spin. We argue that the on-site Hubbard ($U$), nearest-neighbor ($V_1$), and second-neighbor ($V_2$) repulsion when strong enough and of comparable magnitude can produce such an ordered phase as the ground state on a half-filled honeycomb lattice. Furthermore, the application of an in-plane magnetic field, which due to the Zeeman coupling causes the formation of compensated electron and hole Fermi pockets for opposite spin projections, can give rise to excitonic instability toward the ordered state even for weak such repulsions, due to the \emph{Keldysh-Kopaev} mechanism~\cite{keldysh, rice, aleiner, roy-kondo}, with the spin projected  onto the easy-plane perpendicular to the magnetic field. With the Zeeman coupling fully taken into account the massive Dirac Hamiltonian with a half-vortex represents a  generalized Jackiw-Rossi Hamiltonian~\cite{herbut-lu}, and continues to support a single zero energy state. Existence of single zero energy state also provides local expectation values for two competing orders. On the honeycomb lattice they correspond to staggered pattern of electronic density~\cite{semenoff} and topological spin Hall insulator~\cite{haldane}.

\section{Model} 

Although the following discussion is insensitive to the choice of basis, to make our discussion specific to graphene, we work with an appropriate representation for Dirac spinor and $\gamma$ matrices. Linearized dispersion at low energies around two non-equivalent corners of the hexagonal Brillouin zone, chosen here at $\pm \mathbf{K}$, where $\mathbf{K}=\left( 1/ \sqrt{3},1/3\right) 2\pi /a$ and $a$ ($\approx 2.5 \; \mathring{\mbox{A}}$ for graphene) is the lattice spacing, can be captured by an eight component spinor $\Psi=\left( \Psi_\uparrow, \Psi_\downarrow \right)^\top$, with $\Psi^\top_\sigma=$  $\big[ u_\sigma (\mathbf{K}+\mathbf{p})$, $v_\sigma (\mathbf{K}+\mathbf{p})$, $u_\sigma (-\mathbf{K}+\mathbf{p})$, $v_\sigma (-\mathbf{K}+\mathbf{p})\big]$. Fermion annihilation operator on two sublattices A and B of the honeycomb lattice are represented by $u$ and $v$, respectively, and $\sigma=\uparrow, \downarrow$ corresponds to two projections of electron spin along the $z$-direction. When $|\mathbf{p}| \ll |\mathbf{K}|$, the non-interacting tight-binding Hamiltonian with only nearest-neighbor hopping ($t$) assumes a relativistically invariant form $H_0= \sigma_0 \otimes i v \gamma_0 \left( \gamma_1 \hat{p}_1 + \gamma_2 \hat{p}_2 \right)$~\cite{igorisotrpic}, where the Fermi velocity $v\sim t a$, and we set $\hbar=1$, $v=1$. Five mutually anti-commuting $\gamma$ matrices are $\gamma_0 =\sigma_0 \otimes \sigma_3$, $\gamma_1= \sigma_3 \otimes \sigma_2$, $\gamma_2 = \sigma_0 \otimes \sigma_1$, $\gamma_3 = \sigma_1 \otimes \sigma_2$, and $\gamma_5=\sigma_2 \otimes \sigma_2$, where $\sigma_0$ and $\boldsymbol \sigma$ are two dimensional identity and Pauli matrices, respectively.

The low-energy Hamiltonian $H_0$ remains invariant under a global \emph{chiral} $U_c(4)$ rotation, generated by $\left(\sigma_0,\vec{\sigma} \right) \otimes \left( I_4, \gamma_3, \gamma_5, i \gamma_5 \gamma_3\right)$~\cite{igorisotrpic}. In our representation, $\sigma_0 \otimes i \gamma_5 \gamma_3$ stands for the generator of translation~\cite{herbut-juricic-roy}, and $\vec{S}=\vec{\sigma} \otimes I_4$ are the three generators of rotations of electrons spin. Any perturbation, proportional to a generator of the chiral symmetry, reduces the $U_c(4)$ symmetry for $H_0$ down to $U_c(2) \times U_c(2)$. For example, if we introduce a term $H_Z = h \left( \sigma_3 \otimes I_4 \right)$, which one can identify as the Zeeman coupling of the electrons spin to an external magnetic field $B$ applied along the z-direction, the $U_c(2) \times U_c(2)$ symmetry of $H_0+ H_z$ is generated by $\left( \sigma_0, \sigma_3 \right) \otimes \left( I_4, \gamma_3, \gamma_5, i \gamma_5 \gamma_3\right)$. Here $h= g B$, and $g \approx 2$ is the $g$-factor of electrons in graphene.

  Due to vanishing density of states near the band touching points, Dirac fermions are robust against a weak electron-electron interaction. Nevertheless, if the interactions are sufficiently strong, the vacuum can undergo a quantum phase transition into a broken symmetry phase. It is quite natural to expect that the system will minimize the energy in an ordered phase by opening up a \emph{mass gap} at the Dirac points. Order parameter, corresponding to a mass gap, anticommutes with the Dirac Hamiltonian $H_0$. There is therefore a  plethora of broken symmetry phases available to massless  Dirac fermions for condensation~\cite{igorisotrpic, herbut-juricic-roy, raghu, ryu-classification, honerkamp, weeksfranz, herbut-roy-kekuleSC, vozmadiano, gonzalez, juricic-SC}. We will here focus on a particular ordered state which breaks both translational and spin rotation symmetries. Define  the order parameter as $\langle \Psi^\dagger \vec{C} \Psi \rangle $, with the matrix $\vec{C}=(\vec{C}_\perp, C_3)$, where
\begin{equation}
(\vec{C}_\perp, C_3)= \left(\Delta_\perp \vec{\sigma}_\perp, \Delta_3 \sigma_3 \right) \otimes i \gamma_0 \left(  \gamma_3 \cos{\theta_k} + \gamma_5 \sin{\theta_k}\right).
\end{equation}
The amplitudes of the order parameter in the easy-plane and along the magnetic field axis are denoted by $\Delta_\perp$ and $\Delta_3$, respectively, and we will for the moment assume that the angle $\theta_k$ is uniform. In the ordered phase, which in graphene corresponds to the spin-triplet Kekule valence bond solid, the fermion spectrum $\pm \sqrt{\mathbf{p}^2+\Delta^2_\perp+\Delta^2_3}$ is fully gapped. We will discuss its possible microscopic origin shortly.

\section{Half-vortex and fractionalization} 

In the presence of an appropriate chiral symmetry breaking perturbation such as $H_Z$, the easy-axis and easy-plane components of this order parameter affect the energy differently. The spectrum of the mean-field single-particle Hamiltonian in presence of the ordering under consideration, $H_0+H_Z+\vec{C}_\perp+C_3$, is
\begin{equation}
E_{\sigma}(\mathbf{p})=\pm \left[ \left\{ \left(\mathbf{p}^2+ \Delta^2_3 \right)^{1/2} + \sigma h \right\}^2+ \Delta^2_\perp \right]^{1/2},
\end{equation}
for $\sigma=\pm$. Notice that the Zeeman term ($H_Z$) commutes and anticommutes with the easy-axis ($C_3$) and easy-plane ($\vec{C}_\perp$) components of the spin-triplet order ($\vec{C}$), respectively. Hence, the ground state energy of the massive Dirac sea is minimized when the spin component of the triplet order parameter is restricted onto the easy-plane, i.e. for $C_3=0$. We will argue that beside confining the order parameter onto the easy-plane, Zeeman coupling induces such excitonic ordering even for infinitesimally weak repulsive interaction. Assuming that the order parameter is confined to the easy-plane for reasons of energetics, we can cast the above mean-field Hamiltonian into an elegant form
\begin{eqnarray}\label{SPzeeman}
H^\perp_{SP} &=& H_0 + \Delta(r) \left( \sigma_1 \cos{\theta_s} +  \sigma_2 \sin{\theta_s} \right) \nonumber \\
&\otimes& i\gamma_0 \left( \gamma_3 \cos{\theta_k}+ \gamma_5 \sin{\theta_k} \right) + h \left( \sigma_3 \otimes I_4 \right),
\end{eqnarray}
where the second angle $\theta_s$ describes the global direction of the spin in the easy-plane. For generality, we permitted a spatial modulation of the amplitude of the order parameter by taking $\Delta_\perp \to \Delta(r)$, which is assumed to be arbitrary.

To exhibit  the topological properties of this ordered state, we transform $H^\perp_{SP}$ into a block-diagonal form. In the present representation this can be most  easily achieved by exchanging the \emph{second} and the \emph{fourth} $2 \times 2$ blocks. Then, $H^\perp_{SP}$ becomes $H_+ \oplus H_-$, with
\begin{equation}\label{block-diagonal}
H_\pm= H_D + |\Delta(r)| i \gamma_0 \left( \gamma_3 \cos{\theta_\pm} + \gamma_5 \sin{\theta_\pm} \right) \pm h i \gamma_5 \gamma_3,
\end{equation}
where $\theta_\pm= \theta_k \pm \theta_s$ and $H_D= i \gamma_0 (\gamma_1 \hat{p}_1+\gamma_2 \hat{p}_2)$. With the above form of the effective single-particle Hamiltonian one can construct the \emph{half-vortex} topological defect in the ordered phase. If we allow both angles $\theta_k$ and $\theta_s$ to become space dependent and to wind in arbitrary ways from $0$ to $\pi$ around some point, the angle $\theta_+$ will wind by total amount of $2 \pi$ around the same point, whereas the other combination  $\theta_-$ will remain without any winding. In this configuration $H_{+}$ contains a single {\it full} vortex and becomes topologically non-trivial, whereas $H_-$ assumes a topologically trivial background. If, on the other hand, $\theta_k$ and $\theta_s$ wind as before, but with one of them  in the opposite sense, then $H_-$ will host a single vortex, whereas $H_+$ would contain a topologically trivial background. Since the original angles $\theta_k$ and $\theta_s$ wind only by $\pi$, the above defect will be named \emph{half vortex}. Notice that although both angles wind only by $\pm \pi$, the Hamiltonian in Eq.~(\ref{SPzeeman}) or (\ref{block-diagonal}) is nevertheless a single-valued function of coordinate.

 To be specific, let us assume that it is the $H_+$ block that contains a vortex. In the absence of Zeeman coupling $H_+$ is then equivalent  to the Jackiw-Rossi Hamiltonian, and as such it yields a single eigenstate at precise zero energy~\cite{jackiw-rossi, weinberg}. When $h=0$, there exists a unitary operator $\gamma_0$, which anticommutes with $H_+$, ensuring its spectral symmetry, as well as the existence of a single zero energy state. When $h \neq 0$, $\gamma_0$ evidently no longer anticommutes with $H_+$. Nevertheless, even then there exists an anti-unitary operator $A=U K$, where $U$ is a unitary operator and $K$ is the complex conjugation, which still anticommutes with $H_+$~\cite{herbut-lu, goswami-roy}. In our representation $U= i \gamma_2 \gamma_3 = \sigma_1 \otimes \sigma_3$. The spectral symmetry of $H_+$, generated by the \emph{anti-unitary} operator $A$ also ensures the existence of the single zero energy state even in the presence of finite Zeeman coupling. Explicit form of the zero energy state for the simplest choice of
 $\theta_s = \theta_k =\phi/2 $ is
\begin{equation}\label{zeromode}
\Psi_0=\left[ \begin{array}{c}
u_\uparrow (\mathbf K) \\
v_\uparrow (\mathbf K)\\
u_\downarrow (-\mathbf K)\\
v_\downarrow (-\mathbf K)
\end{array}
\right]
= c e^{-i \frac{\pi}{4} - \int^r_0 |\Delta(t)| dt}
\left[ \begin{array}{c}
f(r h) \\
i e^{-i \phi} g(r h)\\
i f(r h)\\
e^{i \phi} g(r h)
\end{array}
\right],
\end{equation}
where $f(rh)=J_0(rh)$, $g(rh)=J_1(rh)$, $\phi$ is the azimuthal angle, and $c$ is the normalization constant. $J_k$s are the Bessel functions of first kind of order $k$. In contrast, $H_-$, containing a topologically trivial background, has no zero modes, and its spectrum is fully gapped.

The existence of a single zero energy state gives a net electrical charge of $\pm e/2$ bound to the half-vortex, depending on whether the zero mode is occupied or vacant, leading to desired fractionalization of electrical charge. In contrast, if we assume that $\theta_s$ is uniform but $\theta_k$ to wind by $2\pi$ \cite{hou-chamon-mudry}, or vice versa \cite{herbut-so3}, there would be {\it two} zero modes, yielding the total bound electrical charge of $e$.

\section{Competing orders} 

The existence of an isolated bound state at zero energy also gives rise to local expectation value of some additional, competing orders~\cite{herbut-so3}.  Any operator that commutes or anti-commutes with the Hamiltonian $H^\perp_{SP}$ in Eq. (\ref{SPzeeman}) leaves the zero energy subspace invariant. In particular, when $h=0$, there are two such operators belonging in the second category, namely $\sigma_0 \otimes \gamma_0$ and $\sigma_3 \otimes i \gamma_1 \gamma_2$. The average of the fermionic bilinear, associated with the former operator
\begin{equation}
\Psi^\dagger \left( \sigma_0 \otimes \gamma_0 \right) \Psi = u^\dagger_\sigma (\vec{p}, \omega) u_\sigma (\vec{p}, \omega) - v^\dagger_\sigma (\vec{p}, \omega) v_\sigma (\vec{p}, \omega),
\end{equation}
can be recognized as the staggered-density-wave in honeycomb lattice. The same quantity for the other operator
\begin{eqnarray}
&& \Psi^\dagger \left( \sigma_3 \otimes i \gamma_1 \gamma_2 \right) \Psi = \sigma \big[ u^\dagger_\sigma (\vec{K}, \omega) u_\sigma (\vec{K}, \omega) \nonumber \\
&-& u^\dagger_\sigma (-\vec{K}, \omega) u_\sigma (-\vec{K}, \omega) \big]- \left[ u \rightarrow v\right],
\end{eqnarray}
represents the $z$-component of the topological quantum spin Hall insulator. Furthermore, it can be shown that these two order parameters acquire their expectation value from the zero energy subspace even when the Zeeman coupling is included. The ground state average of a physical observable $\langle q(\vec{x}) \rangle$, associated with a traceless operator $Q$, can be written as~\cite{herbut-so3, wijewardhana}
\begin{equation}
\langle q(\vec{x}) \rangle = \frac{1}{2} \bigg( \sum_{occupied}- \sum_{empty} \bigg) \psi^\dagger_E Q \psi_E,
\end{equation}
where $\left\{ \psi_E \right\}$ are the eigenstates of a generic Hamiltonian $H$ with energy $E$. If there exist an operator, say $T$, which anti-commutes with $H$, but commutes with $Q$, the above mentioned sum gets restricted to the zero energy subspace. When $h=0$, then we can choose $T=\sigma_0 \otimes i \gamma_1 \gamma_2$ for $Q=\sigma_0 \otimes \gamma_0$, and vice versa. On the other hand, when $h \neq 0$, we cannot find any unitary operator $T$, for either $Q=\sigma_0 \otimes \gamma_0$ or for $\sigma_3 \otimes i \gamma_1 \gamma_2$. However, for both we can choose an antiunitary $T=\sigma_1 \otimes i \gamma_2 \gamma_3 K$.

Consequently, these two order parameters receive the ground state expectation values from the isolated zero mode, which reads as
\begin{equation}
\langle q(r) \rangle =c^{-2} \int^r_0 r \; dr \left[ J^2_0(rh)- J^2_1(rh) \right]\; \exp (-2 \int^r_0 \Delta_t dt),
\end{equation}
for $Q=\sigma_0 \otimes \gamma_0$ as well as $\sigma_3 \otimes i \gamma_1 \gamma_2$.

\section{Graphene}

The ordered phase considered so far on the honeycomb lattice of graphene corresponds to a \emph{spin-triplet Kekule valence bond solid}. In this phase hopping amplitudes between the nearest-neighbor sites acquire a commensurate periodic modulation that is, however, of opposite sign for two projections of electrons spin. We discuss next a possible microscopic origin of this phase.

Spinless fermions in graphene, for example, can spontaneously develop a staggered pattern of charge~\cite{semenoff} or an intra-sublattice circulating current~\cite{haldane}, if the nearest-neighbor ($V_1$) or the next-nearest-neighbor ($V_2$) component of the Coulomb repulsion is sufficiently strong, respectively~\cite{raghu, duric, roy-TI}. When both $V_1$ and $V_2$ are strong and of similar magnitude, however, these two orderings become frustrated, and spinless Dirac fermions can find themselves in the singlet-Kekule phase~\cite{weeksfranz, vozmadiano, gonzalez}. The order parameter for the singlet-Kekule phase reads as $\sigma_0 \otimes i\gamma_0 \left(\gamma_3 \cos{\theta_k}+ \gamma_5 \sin{\theta_k} \right)$~\cite{hou-chamon-mudry, herbut-juricic-roy}. If one neglects the contribution to the ground state energy from the states residing far from the Dirac points, the configurations with different values of $\theta_k$ are \emph{exactly} degenerate, and a \emph{perfect} $U(1)$ symmetry emerges. In the mean field approximation which treats all the quasi-particles as sharp excitations, the contribution from these states \emph{weakly} breaks this degeneracy deep inside the ordered phase, and the configuration with $\theta_k=0$ becomes energetically \emph{slightly} preferred~\cite{weeksfranz, herbut-roy-kekuleSC}. However, as one approaches the transition point from the ordered side, the energy difference among various choices of $\theta_k$ vanishes, restoring a $U(1)$ symmetry at the semimetal-insulator quantum critical point. In addition, upon including the fluctuations, the existence of a massless \emph{Goldstone} mode should enhance the condensation energy gain. Therefore, we believe that lattice-induced $C_{3v}$-symmetric perturbations are irrelevant near the semimetal-insulator transition, and spin-singlet valence bond solid carries a $U(1)$ symmetry, which only gets lifted spontaneously inside the ordered phase.

Once the spin degrees of freedom is restored, the on-site Hubbard interaction ($U$) alone, when strong enough, supports a N\'{e}el antiferromagnetic order~\cite{igorisotrpic, assaad-herbut}. However, various finite range components of the actual long-range Coulomb repulsion are likely to be of comparable strength in graphene~\cite{katsnelson}. Given that $V_1$ and $V_2$, when strong, prefer a singlet valence bond solid~\cite{weeksfranz, vozmadiano, gonzalez}, and that comparable and strong $U$ and $V_2$ stabilize the topological spin Hall insulator~\cite{raghu, roy-TI}, it is conceivable  that the triplet Kekule valence bond solid becomes the preferred ground state when $U$, $V_1$, and $V_2$ are all both strong enough, and of comparable magnitudes.

This possibility notwithstanding, the strength of the actual Coulomb interaction in pristine graphene appears not strong enough to support {\it any} broken symmetry phase. Next we argue that when graphene is placed in a parallel magnetic field, yielding only Zeeman coupling, but no Landau quantization, the proposed ground state can be realized even for sufficiently weak repulsive interactions.

In the presence of the parallel magnetic field, the Zeeman coupling induces compensated electron and hole Fermi pockets for the two projections of the electron spin. Due to the resulting finite density of states, it is well known that a BCS-type instability towards an excitonic ordering sets in even for infinitesimal strength of interactions~\cite{keldysh, aleiner, rice, roy-kondo}. Assuming that the dominant ordering tendency in pristine graphene is towards the formation of triplet Kekule ordering (for comparable $U$, $V_1$, and $V_2$), one expects that its easy-plane version would be the resulting phase in the presence of Zeeman coupling and at infinitesimally weak repulsion.

\section{Conclusion} 

To conclude, we showed that the half vortex in two-dimensional quasi-relativistic systems, in spite of the usual fermion doubling due to the lattice and spin, supports only a single zero mode and thus binds the electrical charge of a $e/2$. In graphene, an example of the ordered phase that allows such a topological defect is the spin-triplet Kekule valence bond solid, with its spin components lying in an easy-plane. We argued that such ordering may be preferred  for weak on-site, nearest-neighbor, and second-neighbor Coulomb repulsion, when these are all of comparable strength, and when graphene is subjected to an in-plane magnetic field.

\acknowledgments  B. R. was supported by NSF-JQI-PFC and LPS-MPO-CMTC. I. F. H acknowledges the support from NSERC of Canada.


\begin{thebibliography}{99}

\bibitem{rajaraman} R.  Rajaraman, \emph{Solitons and Instantons} (North Holland, Amsterdam, 1987).

\bibitem{jackiw-rebbi} R. Jackiw, and C. Rebbi, Phys. Rev. D {\bf 13}, 3398 (1976).

\bibitem{schrieffer} W. P. Su, J. R. Schrieffer, and A. J. Heeger, Phys. Rev. Lett. {\bf 42}, 1698 (1979); Phys. Rev. B {\bf 22}, 2099 (1980).

\bibitem{jackiw-rossi} R. Jackiw, and P. Rossi, Nucl. Phys. B {\bf 190}, 681 (1981).

\bibitem{weinberg} E. J. Weinberg, Phys. Rev. D {\bf 24}, 2669 (1981).

\bibitem{chiken} For an exception, see C.-K. Lu and I. F. Herbut, Phys. Rev. Lett. {\bf 108}, 266402 (2012).


\bibitem{hou-chamon-mudry} C.-Y. Hou, C. Chamon, C. Mudry, Phys. Rev. Lett. {\bf 98}, 186809 (2007).

\bibitem{jackiw-pi} R. Jackiw and S.-Y. Pi, Phys. Rev. Lett. {\bf 98} 266402 (2007).

\bibitem{herbut-so3} I. F. Herbut, Phys. Rev. Lett. {\bf 99}, 206404 (2007).

\bibitem{semenoff-jackiw} C. Chamon, C-Y. Hou, R. Jackiw, C. Mudry, S-Y. Pi, and G. W. Semenoff, Phys. Rev. B {\bf 77}, 235431 (2008).

\bibitem{herbut-cl3} I. F. Herbut, Phys. Rev. Lett. {\bf 104}, 066404 (2010); Phys. Rev. B {\bf 85}, 085304 (2012).

\bibitem{herbut-lu} I. F. Herbut, C.-K. Lu, Phys. Rev. B {\bf 82}, 125402 (2010).

\bibitem{ryu} L. Santos, S. Ryu, C. Chamon, C. Mudry, Phys. Rev. B {\bf 82}, 165101 (2010).

\bibitem{roy-axial} B. Roy, Phys. Rev. B {\bf 85}, 165453 (2012).

\bibitem{roy-smith-kennett} B. Roy, P. M. Smith, and M. P. Kennett, Phys. Rev. B {\bf 85}, 235119 (2012).

\bibitem{doron} D. L. Bergman, Phys. Rev. B {\bf 87}, 035422 (2013).



\bibitem{ninomiya} H.B. Nielsen, and M. Ninomiya, Nucl. Phys. B {\bf 185}, 20 (1981); Phys. Lett. B {\bf 105}, 219 (1981).

\bibitem{keldysh} L. V. Keldysh, Y. V. Kopaev, Sov. Phys. Solid State {\bf 6}, 2219 (1965).

\bibitem{rice} T. M. Rice, Phys. Rev. B {\bf 2}, 3619 (1970).

\bibitem{aleiner} I. L. Aleiner, D. E. Kharzeev, A. M. Tsvelik, Phys. Rev. B {\bf 76}, 195415 (2007).

\bibitem{roy-kondo} B. Roy, J. Hofmann, V. Stanev, J. D. Sau, and V. Galitski, Phys. Rev. B {\bf 92}, 245431 (2015).

\bibitem{semenoff} G. W. Semenoff, Phys. Rev. Lett. {\bf 53}, 2449 (1984).

\bibitem{haldane} F. D. M. Haldane, Physi. Rev. Lett. {\bf 61}, 2015 (1988͒).

\bibitem{igorisotrpic} I. F. Herbut, Phys. Rev. Lett. {\bf 97},146401 (2006).

\bibitem{herbut-juricic-roy} I. F. Herbut, V. Juri\v ci\' c, and B. Roy, Phys. Rev. B {\bf 79}, 085116 (2009).

\bibitem{raghu} S. Raghu, X.-L. Qi, C. Honerkamp, S. C. Zhang, Phys. Rev. Lett. {\bf 100}, 156401 (2008).

\bibitem{honerkamp} C. Honerkamp, Phys. Rev. Lett. {\bf 100}, 146404 (2008).

\bibitem{ryu-classification} S. Ryu, C. Mudry, C-Y. Hou, and C. Chamon, Phys. Rev. B {\bf 80}, 205319 (2009).

\bibitem{weeksfranz} C. Weeks, M. Franz, Phys. Rev. B {\bf 81}, 085105 (2010).

\bibitem{herbut-roy-kekuleSC} B. Roy, I. F. Herbut, Phys. Rev. B {\bf 82}, 035429 (2010).

\bibitem{vozmadiano} E. V. Castro, A. G. Grushin, B. Valenzuela, M. A. H. Vozmadiano, A. Cortijo, and F. de Juan, Phys. Rev. Lett. {\bf 107}, 106402 (2011).

\bibitem{gonzalez} J. Gonzalez, J. High Energy Phys. {\bf 07}, 175 (2013).

\bibitem{juricic-SC}  F. K. Kunst, C. Delerue, C. M. Smith, V. Juricic, Phys. Rev B {\bf 92}, 165423 (2015).

\bibitem{goswami-roy} B. Roy, P. Goswami, Phys. Rev. B {\bf 89}, 144507 (2014).

\bibitem{wijewardhana} G. W. Semenoff, I. A. Shovkovy, and L. C. R. Wijewardhana, Phys. Rev. D {\bf 60}, 105024 (1999).

\bibitem{duric} T. Duri\'{c}, N. Chancellor, and I. F. Herbut, Phys. Rev. B {\bf 89}, 165123 (2014).

\bibitem{roy-TI} See also B. Roy, and I. F. Herbut, Phys. Rev. B {\bf 88}, 045425 (2013).

\bibitem{assaad-herbut} F. F. Assaad, and I. F. Herbut, Phys. Rev. X {\bf 3}, 031010 (2013).

\bibitem{katsnelson} T. O. Wehling, E. \ifmmode \mbox{\c{S}}\else \c{S}\fi{}a\ifmmode \mbox{\c{s}}\else \c{s}\fi{}\ifmmode \imath \else \i \fi{}o\ifmmode \breve{g}\else \u{g}\fi{}lu, C. Friedrich, A. I. Lichtenstein, M. I. Katsnelson, and S. Bl\"ugel, Phys. Rev. Lett. {\bf 106}, 236805 (2011).



\end{thebibliography}
\end{document}